\documentclass[10pt,twocolumn,letterpaper]{article}

\usepackage[pagenumbers]{cvpr} 

\usepackage{graphicx}
\usepackage{amsmath}
\usepackage{amssymb}
\usepackage{booktabs}
\usepackage{multirow}
\usepackage{booktabs}
\usepackage{tabularx}
\usepackage{makecell}
\usepackage{multirow}
\usepackage{makecell}
\usepackage{marvosym}
\usepackage{capt-of}
\usepackage{comment}
\usepackage{multicol}
\usepackage{algorithm}
\usepackage{algcompatible}
\usepackage{algorithmicx}
\usepackage{algpseudocode}
\usepackage{subcaption}
\usepackage[accsupp]{axessibility}

\usepackage{multirow}
\usepackage{colortbl}
\usepackage[dvipsnames]{xcolor}

\newcolumntype{Y}{>{\centering\arraybackslash}X}

\usepackage{stackengine}

\algnewcommand{\LeftComment}[1]{\Statex \(\triangleright\) #1}

\usepackage[dvipsnames]{xcolor}

\definecolor{rred}{RGB}{245, 152, 153}
\definecolor{oorange}{RGB}{253, 205, 154}
\definecolor{yyellow}{RGB}{248,244,140}

\DeclareMathOperator{\diag}{diag}

\definecolor{cvprblue}{rgb}{0.21,0.49,0.74}
\usepackage[pagebackref,breaklinks,colorlinks,citecolor=cvprblue]{hyperref}

\title{Relightable Gaussian Codec Avatars}

\author{
Shunsuke Saito,
Gabriel Schwartz,
Tomas Simon,
Junxuan Li,
Giljoo Nam \\ \\
Codec Avatars Lab, Meta
}

\begin{document}

\twocolumn[{%
\maketitle

\renewcommand\twocolumn[1][]{#1}%
\maketitle
\begin{center}
\centering
\vspace{-20pt}
\captionsetup{type=figure}
\includegraphics[width=\linewidth]{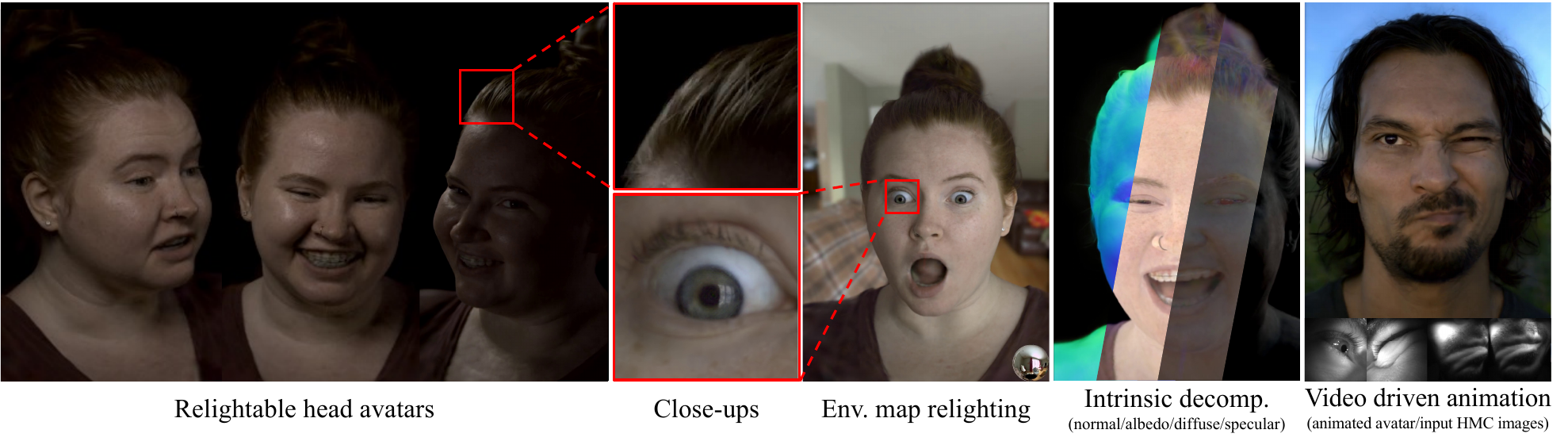} 
\vspace{-22pt}
\caption{\textbf{Relightable Gaussian Codec Avatars.} Our approach enables real-time relighting of human head avatars with all-frequency reflections and detailed hair reconstruction using 3D Gaussians and learnable radiance transfer. Our dynamic avatars can be driven live in real-time from images captured with head mounted cameras (HMC). \url{https://shunsukesaito.github.io/rgca/}}
\label{fig:teaser}
\end{center}
}]

\begin{abstract}
\vspace{-12pt}
The fidelity of relighting is bounded by both geometry and appearance representations.
For geometry, both mesh and volumetric approaches have difficulty modeling intricate structures like 3D hair geometry. 
For appearance, existing relighting models are limited in fidelity and often too slow to render in real-time with high-resolution continuous environments. 
In this work, we present Relightable Gaussian Codec Avatars, a method to build high-fidelity relightable head avatars that can be animated to generate novel expressions. 
Our geometry model based on 3D Gaussians can capture 3D-consistent sub-millimeter details such as hair strands and pores on dynamic face sequences.
To support diverse materials of human heads such as the eyes, skin, and hair in a unified manner, we present a novel relightable appearance model based on learnable radiance transfer. 
Together with global illumination-aware spherical harmonics for the diffuse components, we achieve real-time relighting with all-frequency reflections using spherical Gaussians. 
This appearance model can be efficiently relit under both point light and continuous illumination. 
We further improve the fidelity of eye reflections and enable explicit gaze control by introducing relightable explicit eye models.
Our method outperforms existing approaches without compromising real-time performance. 
We also demonstrate real-time relighting of avatars on a tethered consumer VR headset, showcasing the efficiency and fidelity of our avatars.

\end{abstract}
\vspace{-15pt}
\section{Introduction}
\label{sec:intro}
What makes avatar relighting so challenging? 
Our visual perception is highly sensitive to facial appearance. Convincing the visual system requires modeling each part of the head in sufficient detail that is coherent with an environment, and this synthesis typically needs to be performed in real-time for primary applications of photorealistic avatars including games and telecommunication~\cite{orts2016holoportation,lombardi2018deep}.
Real-time relighting of animatable human heads with convincing details remains an open challenge for three reasons.

The first challenge is that human heads are composed of highly complex and diverse materials that exhibit different properties of scattering and reflectance. 
For example, skin produces intricate reflections due to micro-geometry as well as significant subsurface scattering~\cite{weyrich2006analysis,nagano2015skin}, hair exhibits out-of-plane scattering with multiple reflections due to its translucent fiber structure~\cite{marschner2003light}, and the eyes have multiple layers with highly reflective membranes~\cite{schwartz2020eyes,li2022eyenerf}. 
By and large, there is no single material representation that accurately represents them all, especially in real-time.
Moreover, precise tracking and modeling of the underlying geometry in motion is extremely challenging because deformations do not always contain sufficient visual markers to track.
Finally, the real-time requirement severely limits the algorithmic design. Increase in photorealism traditionally results in an exponential increase in the cost of transporting light and tracking motion.
Our goal is to design a learning framework that builds real-time renderable head avatars with accurate scattering and reflections under illuminations of any frequency. 

Given exhaustive measurements obtained using a light-stage~\cite{ghosh2011multiview,ma2007rapid,debevec2000acquiring}, physically-based rendering methods~\cite{seymour2017meet,weyrich2006analysis} can generalize to novel illuminations. 
However, it remains non-trivial to extend these methods to dynamic performance capture and non-skin parts such as hair and eyeballs. 
Additionally, acquiring sufficiently accurate geometry and material parameters is a laborious process with a significant amount of manual cleanup required~\cite{seymour2017meet}. 

More recently, neural relighting approaches sidestep the need for accurate geometry and material modeling by only modeling the direct relationship between the input (i.e., illumination) and output (i.e., outgoing radiance) with neural networks and approximated geometry using meshes~\cite{bi2020deep}, volumetric primitives~\cite{li2023megane,iwase2023relightablehands,yang2023towards}, and neural fields~\cite{li2022eyenerf,sarkar2023litnerf}.
Typically these models are learned from one-light-at-a-time (OLAT)~\cite{sarkar2023litnerf} or grouped lights~\cite{bi2020deep,iwase2023relightablehands,yang2023towards} controlled by a light-stage, and supporting real-time rendering with continuous illumination requires expensive teacher-student distillation~\cite{bi2020deep,iwase2023relightablehands} or physically-inspired appearance models that explicitly maintain key properties of light transport such as linearity~\cite{li2022eyenerf,yang2023towards}.
Despite promising results, we observe that the existing approaches lead to suboptimal performance due to insufficient expressiveness of both the geometric and appearance representations.
In particular, none of the methods achieve all-frequency reflections on hair and eyes, and submillimeter thin structures such as hair strands are often blurred out or glued into larger blobs, making hair rendering less than photorealistic.

To address the aforementioned issues, we present three contributions: (1) drivable avatars based on 3D Gaussians that can be efficiently rendered with intricate geometric details, (2) a relightable appearance model based on learnable radiance transfer that supports global light transport and all-frequency reflections in real-time, and (3) a relightable explicit eye model that enables disentangled control of gaze from other facial movements as well as all-frequency eye-reflections for the first time in a fully data-driven manner.

\noindent \textbf{3D Gaussian Avatars.}
Our geometric representation is based on 3D Gaussians~\cite{kerbl20233d} that can be rendered in real-time using splatting. 
To achieve a drivable avatar, we decode 3D Gaussians on a shared UV space for a template head using 2D convolutional neural networks.
We encode the driving signals such as facial expressions in a self-supervised manner akin to traditional \emph{codecs}. 
This allows us to track the moving heads in a temporally coherent manner with intricate geometric details such as hair strands. 

\noindent \textbf{Learnable Radiance Transfer.} 
For appearance, inspired by precomputed radiance transfer~\cite{sloan2002precomputed}, we introduce a relightable appearance model based on learnable radiance transfer that consists of diffuse spherical harmonics and specular spherical Gaussians. 
We learn diffuse radiance transfer parameterized by dynamic spherical harmonics coefficients for each 3D Gaussian. This transfer preconvolves visibility and global light transport, including multi-bounce and subsurface scattering.
For specular reflection, we introduce a novel parameterization of spherical Gaussians~\cite{wang2009all,zhang2021physg} with view-dependent visibility that effectively approximates the combined effects of occlusion, Fresnel, and geometric attenuation without explicitly estimating the individual contributions. 
Our specular Gaussian lobe is aligned with the reflection vector and computed using the view direction and per-Gaussian view-dependent normals. 
Most importantly, spherical Gaussians support all-frequency reflection under high-resolution illuminations in real-time.
Both diffuse and specular representations satisfy the linearity of light transport, hence supporting real-time rendering under both point lights and environment illumination without additional training.
In addition, the proposed learnable radiance transfer supports global light transport and all-frequency reflection of the eyes, skin, and hair with the unified representation, significantly simplifying the learning process while achieving extremely high-fidelity relighting.

\noindent \textbf{Relightable Explicit Eye Model.}
To reproduce reflections on the cornea, our relightable Gaussian avatar incorporates an explicit eye model~\cite{schwartz2020eyes} that also enables explicit control of the eyeballs with better disentanglement. 
In addition, our appearance model naturally supports relighting of the eyes with all-frequency reflections, which is crucial for photorealism under natural environments.

We run an evaluation of various pairs of geometry and relightable appearance models in this work and other real-time renderable baseline methods. Our experiments show that the combination of 3D Gaussians with our relighting appearance model outperforms any other combination.

\section{Related Work}
\label{sec:related}

\noindent \textbf{Face Modeling.}
Facial avatar modeling has been an active research topic for over half a century~\cite{parke2008computer}. We refer to \cite{parke2008computer} for a comprehensive overview on research and tools for artist-friendly authoring of 3D facial models. Advancements in image-based 3D reconstruction~\cite{furukawa2009accurate} enabled the precise and more automated acquisition of 3D faces using multi-view capture systems, especially in high-end film production~\cite{zhang2004spacetime,beeler2010high,beeler2011high,bradley2010high,ghosh2011multiview}.
These approaches primarily focus on the facial skin region, and more tailored solutions are required for the reconstruction and modeling of different components such as teeth~\cite{wu2016model}, lips~\cite{garrido2016corrective}, facial hair~\cite{beeler2012coupled}, eyes~\cite{miller2009realistic,berard2016lightweight}, and hair~\cite{paris2008hair,luo2013structure,hu2014robust,nam2019strand}, which are difficult to scale for dynamic and complete head avatars.

More recently, learning-based approaches emerged to holistically represent human heads without requiring precise input geometry~\cite{lombardi2018deep,ma2021pixel,zhao2023havatar,zielonka2023instant}. 
In particular, volumetric representations~\cite{lombardi2019neural,mildenhall2021nerf} show the promise of representing both skin and more complex geometric structures like hair with a single representation~\cite{zhao2023havatar,zielonka2023instant,kirschstein2023nersemble}. 
To enable real-time rendering with volumes, several hybrid approaches are proposed to partition the space for efficient raymarching using mixture of volumetric primitives~\cite{lombardi2021mixture} or tetrahedra~\cite{garbin2022voltemorph}.
Point clouds are also utilized to model head avatars~\cite{zheng2023pointavatar}.
However, we observe that the existing shape representations for real-time renderable avatars struggle with modeling extremely thin structures such as hair strands. To address this limitation, we extend a state-of-the-art efficient scene representation based on 3D Gaussian splatting~\cite{kerbl20233d} to animatable facial avatar modeling. While several works already show dynamic modeling of 3D Gaussians~\cite{luiten2023dynamic,wu20234d}, we are the first to enable the modeling of animatable and (most importantly) relightable 3D Gaussians.

\noindent \textbf{Facial Reflectance Capture.}
In the early 2000s, visual production was a great driver for facial reflectance capture and relighting research for composing actors into virtual environments. 
A seminal work by Debevec \etal~\cite{debevec2000acquiring} demonstrated that one-light-at-a-time (OLAT) captures can be used to obtain reflectance properties and relight faces in novel illuminations by leveraging the linearity of light transport. 
Follow-up work further extended the method to dynamic relighting~\cite{peers2007post}, and fast acquisition of reflectance maps using spherical gradient illuminations~\cite{ma2007rapid,ghosh2011multiview,guo2019relightables}. 
Subsequently, the collection of larger datasets allowed estimating reflectance from a single image using neural networks~\cite{yamaguchi2018high,Li2020LearningFormation,Li2020DynamicFacial,lattas2020avatarme,lattas2021avatarme,papantoniou2023relightify}.  
However, accurate reflectance estimation is typically limited to skin regions because the intricate hair and eye structure make the inverse rendering intractable.
While inverse rendering with various scene representations has also been proposed to estimate spatially-varying BRDFs (SVBRDFs)~\cite{nam2018practical,bi2020deep,zhang2021physg,zhang2021nerfactor,munkberg2022extracting}, it remains a challenge to photorealistically model complete human heads due to the highly complex and diverse material and geometric composition.
The lack of photorealism is also evident in recent relightable head modeling in the wild using simple BRDF and shading models~\cite{zheng2023pointavatar,deng2023lumigan,ranjan2023facelit}.

\noindent \textbf{Neural Relighting.}
Instead of modeling BRDF parameters, learning-based relighting approaches attempt to directly learn relightable appearance from a light-stage capture~\cite{xu2018deep,meka2019deep,gao2020deferred,zhang2021neural,meka2020deep,sarkar2023litnerf,xu2023renerf}. While these approaches show promising relighting for static~\cite{xu2018deep,zhang2021neural,sarkar2023litnerf,xu2023renerf} and dynamic scenes~\cite{meka2019deep,meka2020deep}, they do not support generating novel animations, which is an essential requirement for avatars.
Portrait relighting methods~\cite{sun2019single,tewari2020stylerig,wang2020single,pandey2021total,yeh2022learning} also enable relighting under novel illuminations given a single image. However, they cannot produce novel view synthesis or temporally coherent dynamic relighting. 
Bi \etal~\cite{bi2021deep} propose a neural-rendering method that supports global illumination for animatable facial avatars. To enable real-time rendering with natural environments, they distillate a slow teacher model conditioned with individual point lights into a light-weight student model that can be conditioned with environment maps. This work is later extended to articulated hand modeling~\cite{iwase2023relightablehands}, compositional modeling of heads and eyeglasses~\cite{li2023megane}, and scalable training by eliminating the need of teacher-student distillation~\cite{yang2023towards}.
These relightable avatars take as \textbf{input} the lighting information, which we discover is the main limiting factor for expressiveness in all-frequency relighting. In contrast, inspired by Precomputed Radiance Transfer (PRT)~\cite{sloan2002precomputed,wang2009all}, we propose to integrate a target illumination at the \textbf{output} of our neural decoder, improving quality and simplifying the learning pipeline.
\vspace{5pt}\\ \noindent \textbf{Precomputed Radiance Transfer.}
In computer graphics, rendering a scene with global illumination is an expensive process due to iterative path tracing or multiple bounce computations. To enable real-time rendering with global light transport, Sloan \etal~\cite{sloan2002precomputed} propose to precompute a part of light transport that only depends on intrinsic scene properties, such as geometry and reflectance, and then integrate the precomputed intrinsic factor with an extrinsic illumination at runtime. For fast integration, they utilize spherical harmonics as an angular basis. To overcome the limited frequency band in spherical harmonics, follow-up works introduce wavelets~\cite{ng2003all}, spherical radial basis functions~\cite{tsai2006all}, spherical Gaussians~\cite{green2006view,wang2009all}, anisotropic spherical Gaussians~\cite{xu2013anisotropic}, and neural network-based decompositions~\cite{xu2022lightweight}. 
Similarly, Neural PRT~\cite{rainer2022neural} applies the same principle to screen-space relighting based on neural deferred rendering~\cite{thies2019deferred}. 
Despite many desirable properties, these methods primarily focus on static scenes due to the dependency on knowing geometry and material properties. Unfortunately, we neither know the geometry and material properties for human heads \textit{a priori}, nor are they static. Thus, we propose to \textbf{learn} the intrinsic radiance transfer from dynamic real-data observations without explicitly assuming any material types or underlying geometry.
The closest to our work in terms of the appearance representation is EyeNeRF~\cite{li2022eyenerf}, where they learn view-independent spherical harmonics for diffuse and view-conditioned spherical harmonics for specular components from image observations to build a relightable eye model. However, this appearance model suffers from the limited expressiveness of spherical harmonics for specular reflections. 
Empirically, we find that their proposed model does not generalize well to novel view and light directions. Please refer to Sec.~\ref{sec:experiments} for our analysis.

\begin{figure*}[!t]
\begin{center}
\includegraphics[width=0.95\textwidth]{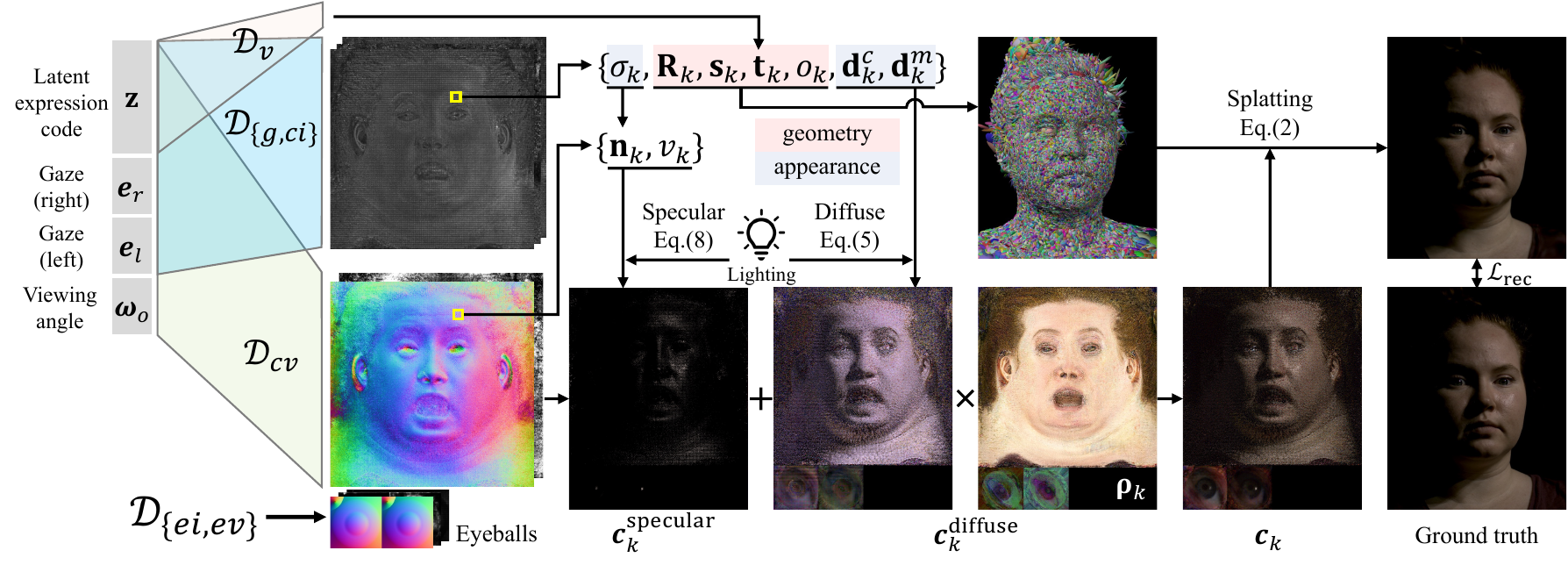} 
\vspace{-5pt}
\caption{\textbf{Overview.} Given an expression latent code $\mathbf{z}$, gaze $\mathbf{e}_{\{l,r\}}$, and view direction $\boldsymbol{\omega}_o$, our model decodes the parameters of 3D Gaussians (rotation $\mathbf{R}_k$, translation $\mathbf{t}_k$, scale $\mathbf{s}_k$, and opacity $o_k$) and learnable radiance transfer functions (colored and monochrome diffuse SH coefficients $\mathbf{d}^c_k$, $\mathbf{d}^m_k$, roughness $\sigma_k$, normal $\mathbf{n}_k$, and visibility $v_k$). We integrate the radiance transfer functions with the input light to compute the final color $\mathbf{c}_k$, which we then render via splatting and supervise in image space. The coarse vertex decoder $\mathcal{D}_{v}$ and geometry decoder $\mathcal{D}_{g}$ are described in Sec.~\ref{sec:geom}, the appearance decoders $\mathcal{D}_{\{ci,cv\}}$ in Sec.~\ref{sec:color}, and eyeball decoders $\mathcal{D}_{\{ei,ev\}}$ in Sec.~\ref{sec:eye}.}
\label{fig:overview}
\vspace{-10pt}
\end{center}
\vspace{-\baselineskip}
\end{figure*}

\section{Method}
\label{sec:method}
In this section, we provide details of the data acquisition process (Sec.~\ref{sec:data}), geometry and appearance representations (Sec.~\ref{sec:geom}--\ref{sec:color}), the relightable explicit eye model (Sec.~\ref{sec:eye}), as well as the training method (Sec.~\ref{sec:train}).

\subsection{Data Acquisition}
\label{sec:data}
We use a setup similar to~\cite{bi2021deep,li2023megane}, where we capture calibrated and synchronized multiview images at $4096 \times 2668$ resolution with 110 cameras and 460 white LED lights at 90 Hz. Each participant is asked to perform a predefined set of various facial expressions, sentences, and gaze motions for about 144,000 frames. To collect diverse illumination patterns while enabling stable facial tracking, we employ time-multiplexed illumination. In particular, the full-on illumination is interleaved at every third frame to allow tracking, and the rest is lit with grouped or random sets of 5 lights. 

As in~\cite{bi2021deep,li2023megane}, we perform a topologically consistent coarse mesh tracking using multi-view full-on images. We further stabilize the head pose using the mode pursuit method of Lamarre \etal~\cite{lamarre2018face}. We also estimate the eye gaze of both eyes using the method described in~\cite{schwartz2020eyes}. The tracked mesh, head pose, unwrapped averaged texture, and gaze are interpolated to adjacent partially lit frames for the following avatar training. 

\subsection{Geometry: 3D Gaussian Avatars}
\label{sec:geom}
The core of our geometric representation is the mixture of 3D anisotropic Gaussians~\cite{kerbl20233d}, which supports representing varying topology and can represent thin volumetric structures. We first review the underlying parameterization and key idea in 3D Gaussian Splatting~\cite{kerbl20233d} and then highlight major changes to enable animatable avatar modeling.

We render avatars as collections of 3D Gaussians, where each Gaussian $\mathbf{g}_k=\{\mathbf{t}_k, \mathbf{R}_k, \mathbf{s}_k, o_k, \mathbf{c}_k\}$ is defined by a translation $\mathbf{t}_k \in \mathbb{R}^3$, a rotation matrix $\mathbf{R}_k \in SO(3)$ parameterized as a quaternion, per-axis scale factors $\mathbf{s}_k \in \mathbb{R}^3$, an opacity value $o_k \in \mathbb{R}$, and a color $\mathbf{c}_k \in \mathbb{R}^3$.
The spatial extent of each Gaussian is defined in 3D by its covariance matrix $\mathbf{\Sigma}_k = \mathbf{R}_k\diag(\mathbf{s}_k)\diag(\mathbf{s}_k)^T\mathbf{R}^T_k$.

This representation allows efficient rendering using the Elliptical Weighted Average (EWA) splatting technique proposed by Zwicker \etal~\cite{zwicker2002ewa} by computing the 2D projection of each 3D Gaussian,
\begin{equation}
    \label{eq:ewa}
    \mathbf{\Sigma}_k^{\prime} = \mathbf{J}\mathbf{V}\mathbf{\Sigma}_k \mathbf{V}^T\mathbf{J}^T,
\end{equation}
where $\mathbf{J} \in \mathbb{R}^{2\times3}$ is the Jacobian of the projective transformation, $\mathbf{V} \in \mathbb{R}^{3\times3}$ the viewing transformation, and $\mathbf{\Sigma}_k^{\prime} \in \mathbb{R}^{2\times2}$ is the covariance matrix of the projected 2D Gaussian (a ``splat''). The final color $\mathbf{C}_p$ at pixel $p$ given $\mathcal{N}$ ordered splats is computed with point-based cumulative volumetric rendering~\cite{kopanas2021point,kopanas2022neural,kerbl20233d} as follows:
\begin{equation}
\label{eq:volrender}
\mathbf{C}_p=\sum_{k \in \mathcal{N}} \mathbf{c}_k \alpha_k \prod_{j=1}^{k-1}\left(1-\alpha_j\right),
\end{equation}
where the transparency $\alpha_k$ is evaluated using the 2D covariance $\mathbf{\Sigma}_k^{\prime}$ and multiplied by the per-Gaussian opacity $o_k$. Please refer to~\cite{kerbl20233d} for more details.

In contrast to 3D Gaussian Splatting~\cite{kerbl20233d}, which focuses on static scene reconstruction, our objective is to build an animatable 3D avatar representation that can span the dynamic facial expressions of a person and also be relit under novel illuminations. 
This necessitates a relightable appearance model to recolor $\mathbf{c}_k$ of all Gaussians as a function of the environmental illumination, allowing for a realistic adaptation of the avatar's appearance under varying lighting conditions. Additionally, it is essential to register the geometry $\{\mathbf{g}_k\}$ for all Gaussians in response to the state of any facial expressions, ensuring that the avatar's expressions remain consistent with the user's actual facial movements. Enabling the encoding and decoding of any facial movements is crucial for animating and driving avatars.

To this end, we parameterize the 3D Gaussians on a shared UV texture map of a coarse template mesh, and decode their transformation and opacity using 2D convolutional neural networks. 
As facial expressions are highly non-linear and non-trivial to precisely define, in the spirit of Lombardi \etal~\cite{lombardi2018deep} and Xu \etal~\cite{xu2023latentavatar}, we employ a conditional variational auto-encoder (CVAE)~\cite{kingma2013auto} to learn the latent distribution of facial expressions from the data. Given the eye gaze directions of both eyes $\mathbf{e}_{\{l,r\}} \in \mathbb{R}^3$ in head-centric coordinates, coarse tracked mesh vertices $\mathbf{V}$, an unwrapped averaged texture $\mathbf{T}$, our encoder $\mathcal{E}$ and geometry decoder $\mathcal{D}_g$ are defined as:
\begin{equation}
\begin{gathered}
\label{eq:geo_encdec}
\mathbf{\mu}_e, \mathbf{\sigma}_e = \mathcal{E}(\mathbf{V}, \mathbf{T}; \Theta_e), \\
\{\mathbf{\delta t}_k, \mathbf{R}_k, \mathbf{s}_k, o_k\}_{k=1}^{M} = \mathcal{D}_g(\mathbf{z}, \mathbf{e}_{\{l,r\}}; \Theta_g), 
\end{gathered}
\end{equation}
where $\Theta_e$ and $\Theta_g$ are the learnable parameters for the encoder and decoder respectively, $M$ is the total number of Gaussians, and $\mathbf{\sigma}_e$ and $\mathbf{\mu}_e$ are the mean and standard deviation of a normal distribution $\mathbf{z} \sim \mathcal{N}(\mathbf{\mu}_e, \mathbf{\sigma}_e)$. The sampled latent vector $\mathbf{z} \in \mathbb{R}^{256}$ is computed using the reparameterization trick proposed by Kingma and Welling~\cite{kingma2013auto}. We also decode mesh vertices from $\mathbf{z}$ such that we can animate the avatars from a headset~\cite{wei2019vr} or latent-space manipulation~\cite{abdal2019image2stylegan}:
\begin{equation}
\mathbf{V}^{\prime} = \mathcal{D}_v(\mathbf{z}; \Theta_v).
\end{equation}
Note that while we directly infer the rotation $\mathbf{R}$ and scale $\mathbf{s}$, we use the coarse geometry $g$ as guidance to avoid poor local minima under large motions. The final Gaussian position $\mathbf{t}_k$ is computed as $\mathbf{t}_k = \hat{\mathbf{t}}_k + \mathbf{\delta t}_k$, where $\hat{\mathbf{t}}_k$ is the interpolated coarse mesh position of the corresponding UV-coordinates using barycentric interpolation of the vertices in $\mathbf{V}^{\prime}$. To assign UV-coordinates to Gaussians, we map one Gaussian to each texel in the UV map.

While the aforementioned parameterization is similar to the Mixture of Volumetric Primitives (MVP)~\cite{lombardi2021mixture}, where a collection of voxel grids is anchored on a template mesh and used as the renderable primitive, there are two important differences: (1) Unlike MVP, which requires raymarching, the 3D Gaussians can be more efficiently rendered using splatting~\cite{kerbl20233d,zwicker2002ewa}. (2) Additionally, the Gaussians have a greater ability to recreate thin structures, yielding sharper appearance for hair, which we show in Sec.~\ref{sec:experiments}.

\subsection{Appearance: Learnable Radiance Transfer}
\label{sec:color}
An appearance model for faces must accurately model a wide range of light transport effects, including subsurface scattering in skin, and specular reflections on the skin, eyes, as well as multi-bounced scattering on the hair. 
As discussed in early works~\cite{sloan2002precomputed,wang2009all}, while a diffuse transfer operator is a low-pass filtering of incident illumination, a specular transfer operator requires the ability to represent all-frequency information for mirror-like reflections. 
To effectively allocate the capacity of the network to each component, we decompose the final color $\mathbf{c}_k$ of each 3D Gaussian into a view-independent diffuse term $\mathbf{c}^{\textrm{diffuse}}_{k}$, and a view-dependent specular term $\mathbf{c}^{\textrm{specular}}_{k}(\boldsymbol{\omega}_o)$, such that $\mathbf{c}_k = \mathbf{c}^{\textrm{diffuse}}_{k} + \mathbf{c}^{\textrm{specular}}_{k}(\boldsymbol{\omega}_o)$, where $\boldsymbol{\omega}_o$ is the viewing direction. 

\noindent \textbf{Diffuse Color.}
Our diffuse term is based on spherical harmonics (SH), and incorporates global light transport effects including occlusion, subsurface scattering, and multi-bounce scattering~\cite{sloan2002precomputed}. The diffuse color contribution $\mathbf{c}^{\textrm{diffuse}}_{k}$ is computed by the spherical integration of an (extrinsic) incident illumination, $\mathbf{L}(\cdot)$, with an intrinsic function that models the radiance transfer, $\mathbf{d}(\cdot)$. By representing both functions in the SH basis, this can be efficiently computed as a dot product of the coefficient vectors due to orthonormality of the basis:
\begin{equation}
\label{eq:diff}
\mathbf{c}^{\textrm{diffuse}}_{k} = \boldsymbol{\rho}_k \odot\int_{\mathbb{S}^2} \mathbf{L}(\boldsymbol{\omega}_i) \odot\mathbf{d}_k(\boldsymbol{\omega}_i) \mathrm{d} \boldsymbol{\omega}_i = \boldsymbol{\rho}_k\odot\sum_{i=1}^{(n+1)^2}{\mathbf{L}_{i}\odot\mathbf{d}_{k}^{i}},
\end{equation}
where $\mathbf{L} = \{\mathbf{L}_{i}\}$ and $\mathbf{d}_{k} = \{\mathbf{d}^{i}_{k}\}$ are the $n$-th order SH coefficients of the incident light and the intrinsic radiance transfer function, with $\mathbf{d}^{i}_{k}\in\mathbb{R}^3$, and $\boldsymbol{\rho}_k\in\mathbb{R}^3$ a learnable albedo which we statically define on each Gaussian to encourage temporally consistent reconstructions.
While diffuse light transport is a low-pass filter that requires only 2nd or 3rd order SH~\cite{ramamoorthi2001efficient}, this is not sufficient to represent shadows. To enable higher-frequency shadows while decoding a manageable number of coefficients that can fit on consumer GPU memory, we propose to decode RGB intrinsic SH coefficients $\mathbf{d}^{\textrm{c}}_{k}$ up to the 3rd order, and only monochrome intrinsic SH coefficients $\mathbf{d}^{\textrm{m}}_{k}$ from 4-th to 8-th order. 

\noindent \textbf{Specular Reflection.}
To achieve sharp, mirror-like reflections, for the view-dependent specular term we use a spherical Gaussian (SG) as an angular basis.
In particular, we propose a normalized, angle-based spherical Gaussian:
\begin{equation}
\label{eq:sg}
  G_{s}(\mathbf{p}; \mathbf{q}, \sigma) = Ce^{-\frac{1}{2}(\frac{\arccos(\mathbf{p}\cdot \mathbf{q})}{\sigma})^2}, 
\end{equation}
where $\sigma \in \mathbb{R}^+$ is the standard deviation of angular decay, $\mathbf{q} \in \mathbb{S}^2$ is the lobe axis, $\mathbf{p} \in \mathbb{S}^2$ is the direction of evaluation, and $C=1/(\sqrt{2}\pi^{2/3}\sigma)$ is a normalization factor to preserve the integral of the Gaussian.

Note that this parameterization is different from the more widely used parameterization $G(\mathbf{p} ; \mathbf{q}, \lambda, \mu)=\mu e^{\lambda(\mathbf{p} \cdot \mathbf{q}-1)}$~\cite{wang2009all,zhang2021physg}, but we observed that this choice often fails to model highly reflective surfaces, such as the cornea. 

As the majority of specular BRDFs have a lobe that is axis-aligned with a particular reflection direction, we compute a reflection vector as the lobe axis:
\begin{equation}
\label{eq:ref}
\mathbf{q}_k = 2(\boldsymbol{\omega}^o_k \cdot \mathbf{n}_k)\mathbf{n}_k - \boldsymbol{\omega}^o_k,
\end{equation}
where $\boldsymbol{\omega}^o_k$ is the viewing direction evaluated at the Gaussian center, and $\mathbf{n}_k$ is a normal direction computed for each Gaussian, and $\mathbf{q}_k$ is the lobe axis.
Our final color for the specular term of each 3D Gaussian is defined as follows:
\begin{equation}
\label{eq:spec}
\mathbf{c}^{\textrm{specular}}_{k}(\boldsymbol{\omega}_o) = v_k(\boldsymbol{\omega}_o) \int_{\mathbb{S}^2} \mathbf{L}(\boldsymbol{\omega}_i)G_s(\boldsymbol{\omega}_i; \mathbf{q}_k, \sigma_{k}) \mathrm{d} \boldsymbol{\omega}_i,
\end{equation}
where $v_k(\boldsymbol{\omega}_o) \in(0,1)$ is a learnable view-dependent visibility term that accounts for Fresnel effects and geometric occlusion integrated within the BRDF lobe. 
Please refer to Appendix~\ref{sec:discussion} for the connection to the rendering equation~\cite{kajiya1986rendering}. For point lights, we use a Dirac delta function multiplied by a light intensity as the incident light $\mathbf{L}(\boldsymbol{\omega}_i)$ for fast evaluation, but an SG parameterization with closed form integrals~\cite{wang2009all} would also be possible for directional area lights. Importantly, Eq.~\ref{eq:spec} can be efficiently evaluated with negligible overhead for all-frequency continuous illumination by prefiltering the environment maps~\cite{kautz2000unified,mcallister2002efficient} as demonstrated in Fig.~\ref{fig:teaser}. This requires only a single mipmap texture look up per 3D Gaussian, which is a critical property for real-time relighting with all-frequency reflections. 

Although the aforementioned formulation works well for surfaces, we use 3D Gaussians to also represent thin fiber-like structures such as hair strands, where Eq.~\ref{eq:spec} is incorrect if a viewer rotates along a tangent vector of each fiber~\cite{kajiya1989rendering}.
To support specular reflection of both surfaces and fibers in a unified manner, we propose to learn a view-conditioned surface normal. 
While the learned normal can remain constant under view changes for surface regions, the normal can rotate along the tangent axis based on the view direction for fibers. 
This way, each 3D Gaussian can flexibly choose its underlying reflection behavior without requiring predefining it a priori.
Our learnable view-dependent normal also supports the case where BRDF lobes are not exactly aligned with the surface normal~\cite{marschner2003light} by adjusting the normal orientation.

\noindent \textbf{Decoder.}
Similar to the geometric decoder, we decode radiance transfer parameters using a view-independent decoder $\mathcal{D}_{ci}$ and a view-dependent decoder $\mathcal{D}_{cv}$ as follows:
\begin{equation}
\begin{gathered}
\{\mathbf{d}^{\textrm{c}}_{k}, \mathbf{d}^{\textrm{m}}_{k}, \sigma_{k}\}_{k=1}^{M} = \mathcal{D}_{ci}(\mathbf{z}, \mathbf{e}_{\{l,r\}}; \Theta_{ci}), \\
\{\mathbf{\delta n}_k, v_k\}_{k=1}^{M} = \mathcal{D}_{cv}(\mathbf{z}, \mathbf{e}_{\{l,r\}}, \boldsymbol{\omega}_o; \Theta_{cv}),
\end{gathered}
\label{eq:spec_dec}
\end{equation}
where $\Theta_{ci}$ and $\Theta_{cv}$ are the learnable parameters of each decoder, and the normal residual $\mathbf{\delta n}_k$ is added to the barycentric interpolated coarse mesh normal $\hat{\mathbf{n}}_k$ to obtain the final normal $\mathbf{n}_k$ as follows: $\mathbf{n}_k = (\hat{\mathbf{n}}_k + \mathbf{\delta n}_k)/\| \hat{\mathbf{n}}_k + \mathbf{\delta n}_k\|$. In practice, since $\mathcal{D}_g$ in Eq.~\ref{eq:geo_encdec} and $\mathcal{D}_{ci}$ in Eq.~\ref{eq:spec_dec} take the same input and produce per-Gaussian values, we model them using a single decoder.

\subsection{Relightable Explicit Eye Model}
\label{sec:eye}
To enable better disentanglement and high-fidelity eye relighting, we use an explicit eye model proposed by Schwartz \etal~\cite{schwartz2020eyes} as the underlying geometric representation of eyes. 
In particular, we parameterize eyeballs as the smooth blending of two spheres; one accounts for the eyeball and the other for the cornea. They are explicitly rotated based on a gaze direction. Each eyeball is parameterized by $\mathbf{E}=\{r_e, r_c, d, c_e\}$ with the radii of the eyeball $r_e$ and cornea $r_c$, the offset $d$ along the optical axis from the center of the eyeball to the center of the cornea sphere, and the center of the eyeball $c_e$ relative to the head in a canonical space. We first optimize $E$ following the optimization presented in Schwartz \etal~\cite{schwartz2020eyes}, and then jointly refine it end-to-end with the other parameters of an avatar. 

While we use the same geometric and appearance representations for eyeballs as the rest of the head, we observe that additional modification is required to enable high-fidelity eye relighting. 
Since the cornea exhibits mirror-like reflections, the discrete point lights of our capture system create reflections that span only a few pixels (often a single Bayer cell) and saturate the sensor. The remaining region has nearly zero contribution. Due to this highly discrete signal, 3D Gaussians quickly fall into poor local minima and fail to correctly model eye glints if we independently optimize the position and surface normal of each Gaussian. Therefore, we freeze the position of Gaussians on the surface of the eyeballs and fix their normals to be the surface normals of the eyeball mesh. In addition, the iris is observed only through the transparent cornea, with significant refraction. To support refraction with the same underlying appearance representation, we use a view-conditioned albedo for the eyes. This effectively allows the eye diffuse color to account for refraction based on the input viewpoint.

To incorporate these modifications, our geometry and appearance eye decoders for each eye are defined as follows:
\begin{equation}
\begin{gathered}
\label{eq:spec_dec2}
\{\mathbf{R}_k, \mathbf{s}_k, o_k,\mathbf{d}^{\textrm{c}}_{k}, \mathbf{d}^{\textrm{m}}_{k}, \sigma_{k}\}_{k=1}^{M_e} = \mathcal{D}_{ei}( \mathbf{e}, \mathbf{h}_p, \mathbf{h}_r; \Theta_{ei}), \\
\{\boldsymbol{\rho}_k,v_k\}_{k=1}^{M_e} = \mathcal{D}_{ev}(\mathbf{e}, \mathbf{h}_p, \mathbf{h}_r, \boldsymbol{\omega}_o; \Theta_{ev}),
\end{gathered}
\end{equation}
where $\mathcal{D}_{ei}$ and $\mathcal{D}_{ev}$ are view-independent and view-conditioned eye decoders with parameters $\Theta_{ei}$ and $\Theta_{ev}$ respectively, and the relative head position $\mathbf{h}_p \in \mathbb{R}^3$ and rotation $\mathbf{h}_r \in SO(3)$ and the gaze $\mathbf{e}$ are used to absorb tracking errors.

\subsection{Training}
\label{sec:train}

\begin{table}[t!]
    \centering
    \begin{footnotesize}
    \caption{\textbf{Comparison on held-out segments}. The top three techniques are highlighted in \textcolor{rred}{red}, \textcolor{oorange}{orange}, and \textcolor{yyellow}{yellow}, respectively.}
    \vspace{-5pt}
    \label{tab:abl-feature}
    \setlength{\tabcolsep}{2mm}{
    \renewcommand\arraystretch{0.975}
    \resizebox{1.0\columnwidth}{!}{
    \begin{tabular}{c|c|c|ccc}
    \toprule
    \multirow{2}{*}{ } &\multirow{2}{*}{Geometry} & \multirow{2}{*}{Appearance}& \multicolumn{3}{c}
    {Metrics}\\
    & & & \footnotesize{PSNR\;$\uparrow$} & \footnotesize{SSIM\;$\uparrow$} & 
    \footnotesize{LPIPS\;$\downarrow$} \\
    \midrule
    A & \multirow{2}{*}{Ours w/ EEM} &EyeNeRF~\cite{li2022eyenerf}&34.550&0.939&0.115\\
    B & &Ours&\cellcolor{oorange}36.501&\cellcolor{rred}0.943&\cellcolor{rred}0.110\\
    \midrule
    C & \multirow{3}{*}{Ours} &EyeNeRF~\cite{li2022eyenerf}&35.110&0.938&\cellcolor{yyellow}0.113\\
    D & &Linear~\cite{yang2023towards}&33.831&0.936&0.184\\
    E& &Ours&\cellcolor{rred}36.529&\cellcolor{oorange}0.943&\cellcolor{oorange}0.111\\
     \midrule
    F & \multirow{3}{*}{MVP~\cite{lombardi2021mixture}} &EyeNeRF~\cite{li2022eyenerf}&27.594&0.922&0.151\\
    G& &Linear~\cite{yang2023towards}&\cellcolor{yyellow}36.294&0.942&0.140\\
    H & &Ours &35.789&\cellcolor{yyellow}0.943&0.134\\
    \bottomrule
    \end{tabular}}}
    \vspace{-10pt}
    \end{footnotesize}
\end{table}

Given multiview video data of a person illuminated with known point light patterns, 
we jointly optimize all trainable network parameters $\Theta$, the static albedo $\boldsymbol{\rho}$, and the eyeball parameters $\mathbf{E}_{\{l,r\}}$ with the following loss function $\mathcal{L}$:
\begin{equation}
\label{eq:loss_all}
\mathcal{L} = \mathcal{L}_{\mathrm{rec}} + \mathcal{L}_{\mathrm{reg}} + \lambda_{\mathrm{kl}} \mathcal{L}_{\mathrm{kl}},
\end{equation}
where $\mathcal{L}_{\mathrm{kl}}$ is the KL-divergence loss on our encoder outputs.
The reconstruction loss $\mathcal{L}_{\mathrm{rec}}$ consists of L1 and D-SSIM on the rendered image as in the original 3DGS paper~\cite{kerbl20233d,wang2004imagequality} as well as L2 loss on the coarse geometry $V'$ as follows: 
\begin{equation}
\label{eq:loss_recon}
\mathcal{L}_{\mathrm{rec}} = \lambda_{\mathrm{l1}} \mathcal{L}_{\mathrm{l1}} + \lambda_{\mathrm{ssim}} \mathcal{L}_{\mathrm{ssim}} + \lambda_{\mathrm{geo}} \mathcal{L}_{\mathrm{geo}}.
\end{equation}
The regularization loss is defined as:
\begin{equation}
\label{eq:loss_reg1}
\mathcal{L}_{\mathrm{reg}} = \lambda_{\mathrm{s}} \mathcal{L}_{\mathrm{s}} + \lambda_{\mathrm{c-}} \mathcal{L}_{\mathrm{c-}} + \lambda_{\mathrm{es}} \mathcal{L}_{\mathrm{es}} + \lambda_{\mathrm{ev}} \mathcal{L}_{\mathrm{ev}} + \lambda_{\mathrm{eo}} L_{\mathrm{eo}}.
\end{equation}
The scale regularization term $\mathcal{L}_{\mathrm{s}}$ encourages the Gaussian scale $\{\mathbf{s}_k\}$ to stay within a reasonable range as follows:
\begin{equation}
\begin{aligned}
\label{eq:loss_reg2}
\mathcal{L}_{\mathrm{s}} = \mathrm{mean}(l_{\mathrm{s}}), \, l_{\mathrm{s}} &= 
\begin{cases}
1/\max(s,10^{-7})  & \text{if } s < 0.1 \\
(s - 10.0)^2 & \text{if } s > 10.0\\
0 & \text{otherwise},
\end{cases}
\end{aligned}
\end{equation}
where $s$ denotes the scale value of each axis in each Gaussian, and $\mathrm{mean}(\cdot)$ is the average operation across all dimensions. The negative color loss $\mathcal{L}_{\mathrm{c-}}$ penalizes negative color in the diffuse term as SH can yield negative values:
\begin{equation}
\label{eq:loss_neg}
\mathcal{L}_{\mathrm{c-}} = \mathrm{mean}(l_{\mathrm{c-}}), \, l_{\mathrm{c-}} = \min(\mathbf{c}^{\textrm{diffuse}}_{k}, 0)^2.
\end{equation}
Also, three regularization terms are used to prevent eye Gaussians from becoming transparent as follows:
\begin{equation}
\begin{gathered}
\label{eq:loss_eye}
\mathcal{L}_{\mathrm{es}} = \mathrm{mean}(l_{\mathrm{es}}), \, l_{\mathrm{es}} = \max(s - 0.1, 0)^2, \\
\mathcal{L}_{\mathrm{eo}} = \mathrm{mean}(l_{\mathrm{eo}}), \, l_{\mathrm{eo}} = (1 -o_{k})^2, \\
\mathcal{L}_{\mathrm{ev}} = \mathrm{mean}(l_{\mathrm{ev}}), \, l_{\mathrm{ev}} = (1-v_{k})^2.
\end{gathered}
\end{equation}
The relative weights are $\lambda_{\mathrm{geo}}=\lambda_{\mathrm{l1}}=10$, $\lambda_{\mathrm{ssim}}=0.2$, $\lambda_{\mathrm{s}}=\lambda_{\mathrm{c-}}=\lambda_{\mathrm{es}}=1.0\times10^{-2}$, $\lambda_{\mathrm{eo}}=\lambda_{\mathrm{ev}}=1.0\times 10^{-4}$, and $\lambda_{\mathrm{kl}}=2.0\times 10^{-3}$.
We use the Adam optimizer~\cite{kingma2014adam} with a learning rate of $0.0005$. We train our model on 4 NVIDIA A100 GPUs with a batch size of 16 for 200k iterations. Please refer to Appendix~\ref{sec:net_arch} for network architecture.

\section{Experiments}
\label{sec:experiments}

\paragraph{Evaluation Protocol.}
We selected three subjects for quantitative evaluations and three more subjects for qualitative results with diverse races, genders, and hairstyles. Our evaluation included around 9,000 conversational expression frames and about 100 \textit{disgust} expression frames not seen during training. 
We also exclude 10 unique frontally-biased light patterns entirely from the training. This corresponds to approximately 1800 out of 144,000 frames.
We report PSNR, SSIM, and LPIPS~\cite{zhang2018unreasonable} on images masked with the face region to avoid influence from the background.

\subsection{Qualitative Results}

Fig.~\ref{fig:teaser} shows that our reconstructed avatars generalize to novel views, expressions, and illuminations including point lights and high-resolution environment maps. Notice the mirror-like reflections in the eyes that faithfully represent the environment without losing high-frequency details. As our model is drivable and supports real-time relighting, real-time driving from a headset is also possible~\cite{wei2019vr}.

\begin{figure}[!t]
\begin{center}
\includegraphics[width=\linewidth]{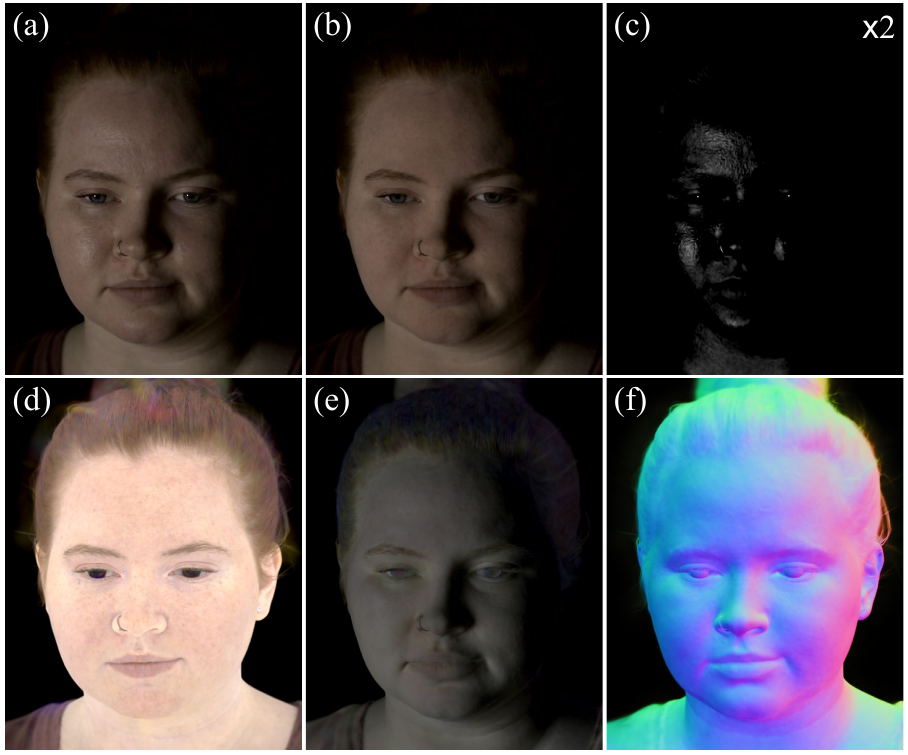} 
\vspace{-15pt}
\caption{\textbf{Intrinsics decomposition}. The full render (a) is produced by addition of a diffuse (b) and a specular component (c) (intensity multiplied by 2 for clarity). The diffuse component is obtained by multiplying a learned albedo (d) with shading computed by SH-based radiance transfer (e). The specular lobes direction is computed using a per-Gaussian normal (f). }
\label{fig:decomp}
\vspace{-25pt}
\end{center}
\end{figure}

While this is not our primary goal, as a bi-product, our approach estimates intrinsic properties of reflectance including albedo, geometry, surface normal, multi-bounce scattering, and specular components in a self-supervised manner.
As shown in Fig.~\ref{fig:decomp}, our method enables 3D consistent and high-fidelity intrinsics decomposition.

\begin{table}[t]
    \centering
    \begin{footnotesize}
    \caption{\textbf{Comparison on held-out lights}. The top three techniques are highlighted in \textcolor{rred}{red}, \textcolor{oorange}{orange}, and \textcolor{yyellow}{yellow}, respectively.}
    \vspace{-5pt}
    \label{tab:abl-feature2}
    \setlength{\tabcolsep}{2mm}{
    \renewcommand\arraystretch{0.975}
    \resizebox{1.0\columnwidth}{!}{
    \begin{tabular}{c|c|c|ccc}
    \toprule
    \multirow{2}{*}{ } &\multirow{2}{*}{Geometry} & \multirow{2}{*}{Appearance}& \multicolumn{3}{c}
    {Metrics}\\
    & & & \footnotesize{PSNR\;$\uparrow$} & \footnotesize{SSIM\;$\uparrow$} &
    \footnotesize{LPIPS\;$\downarrow$} \\
    \midrule
    A& \multirow{2}{*}{Ours w/ EEM} &EyeNeRF~\cite{li2022eyenerf}&30.7976&0.828&\cellcolor{yyellow}0.162\\
    B & &Ours&\cellcolor{rred}34.042&\cellcolor{yyellow}0.858&\cellcolor{rred}0.148\\
    \midrule
    C & \multirow{3}{*}{Ours} &EyeNeRF~\cite{li2022eyenerf}&30.836&0.815&0.163\\
    D & &Linear~\cite{yang2023towards}&32.829&\cellcolor{oorange}0.870&0.202\\
    E & &Ours&\cellcolor{oorange}33.845&0.831&\cellcolor{oorange}0.148\\
     \midrule
    F & \multirow{3}{*}{MVP~\cite{lombardi2021mixture}}  &EyeNeRF~\cite{li2022eyenerf}&28.030&0.812&0.210\\
    G & &Linear~\cite{yang2023towards}&33.444&0.726&0.192\\
    H &  &Ours&\cellcolor{yyellow}33.778&\cellcolor{rred}0.877&0.168\\
    \bottomrule
    \end{tabular}}}
    \vspace{-10pt}
    \end{footnotesize}
\end{table}

\subsection{Discussion}\label{sec:comparison}
\begin{figure}[!ht]
\begin{center}
\includegraphics[width=\linewidth]{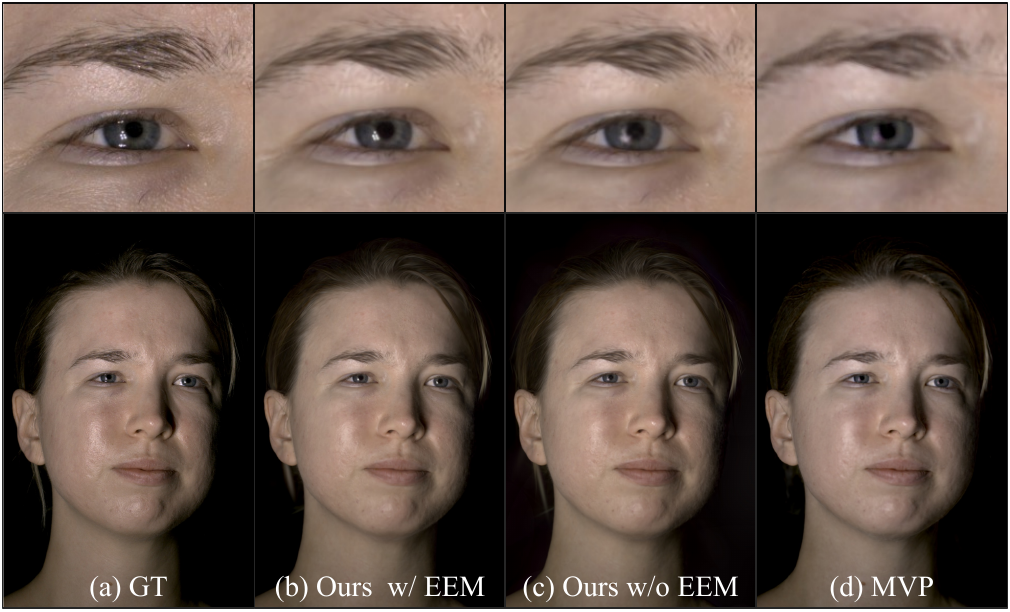} 
\vspace{-15pt}
\caption{\textbf{Geometric representation comparison}. Compared to a held out frame, (a), our Gaussian splatting decoded geometry (b,c) shows improved resolution over MVP~\cite{lombardi2021mixture} (d), especially in fine details like eyelashes and pores. The explicit eyeball model (b) additionally improves realism in eye glints. All methods use the appearance model described in Sec.~\ref{sec:color}.}
\label{fig:Geo_compare}
\vspace{-15pt}
\end{center}
\end{figure}

\noindent \textbf{Geometric Representation.}
We evaluate the geometry component by comparing three variations: our proposed method, our method excluding the explicit eye model (EEM)~\cite{schwartz2020eyes}, and voxel-based primitives~\cite{lombardi2021mixture}. 
For fair comparison, we use the same appearance model and only change the geometric representation (Tab.~\ref{tab:abl-feature} and Tab.~\ref{tab:abl-feature2}  B, D, H).
Fig.~\ref{fig:Geo_compare} clearly demonstrates that our geometry based on 3D Gaussians can better model skin details and hair strands than MVP. Further, our full model, when combined with an EEM, achieves convincing eye glints.

\begin{figure}[!ht]
\begin{center}
\includegraphics[width=\linewidth]{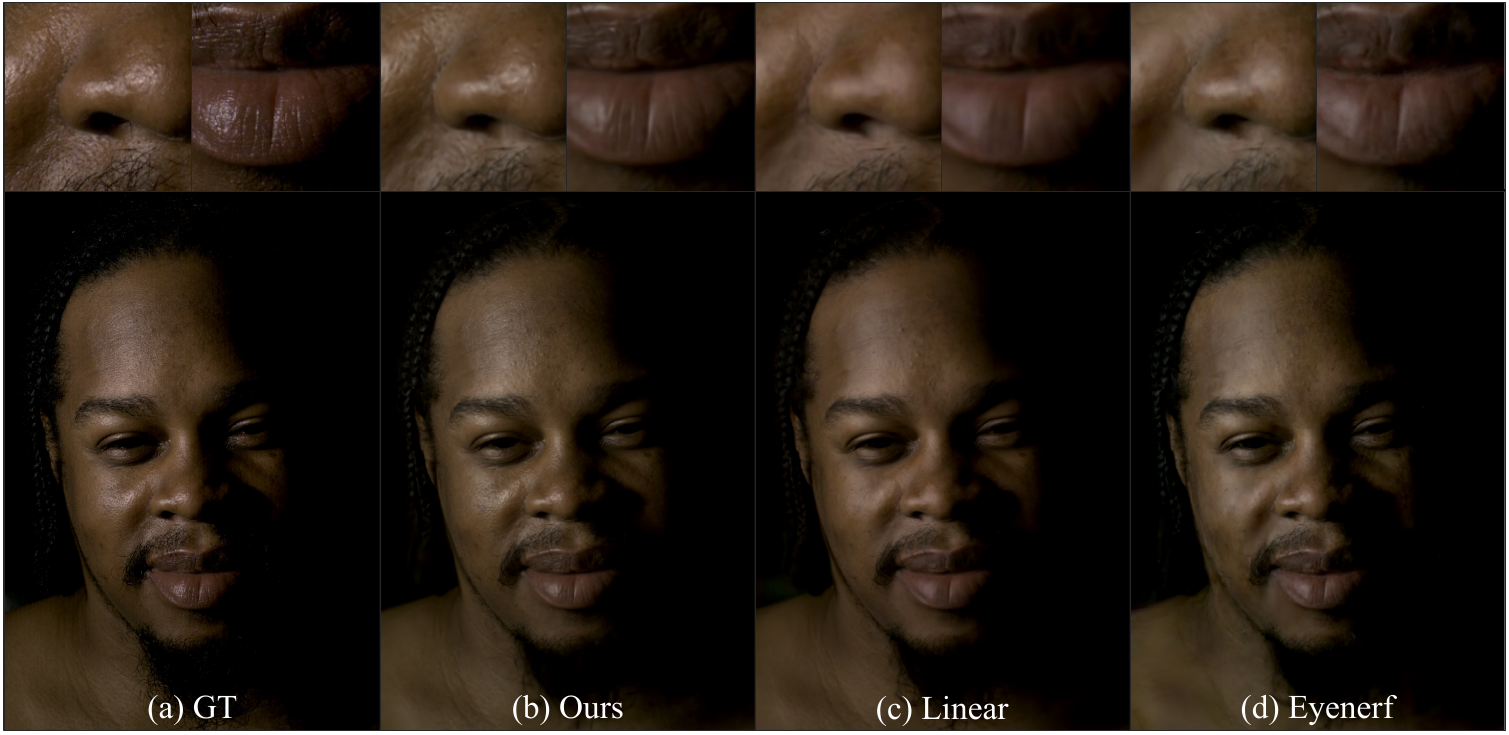} 
\vspace{-15pt}
\caption{\textbf{Appearance representation comparison}. Compared to a held out frame (a), our appearance model (Sec.~\ref{sec:color}) shows sharper pore-level specularities than methods using only a linear neural network~\cite{yang2023towards} or the spherical harmonics-only method ``Eyenerf''~\cite{li2022eyenerf}. All methods use the geometric representation described in Sec.~\ref{sec:geom} (without explicit eyeballs.)}
\label{fig:App_compare}
\vspace{-25pt}
\end{center}
\end{figure}

\noindent \textbf{Appearance Representation.}
For appearance representation, we compare our appearance model with existing relightable appearance representations that support rendering with environment maps in real-time. The model presented by Yang \etal~\cite{yang2023towards} is a linear neural network that explicitly retains the linearity of light transport (denoted as Linear), demonstrating superior performance than a previous state-of-the-art method~\cite{bi2021deep}. For this reason, we omit the comparison with~\cite{bi2021deep}.
To evaluate the effectiveness of our specular reflection model, we also replace our specular component with view-dependent spherical harmonics proposed by EyeNeRF~\cite{li2022eyenerf}.
Tab.~\ref{tab:abl-feature} and Tab.~\ref{tab:abl-feature2} C, D, E show that our appearance representation outperforms existing appearance models in most of the metrics.
As shown in Fig.~\ref{fig:App_compare}, while the linear model produces correct overall color, the relighting result is blurry and lacks high-frequency details. This is primarily limited by the bottleneck lighting representation. The view-dependent spherical harmonics in EyeNeRF shows more detailed reflections, but its expressiveness is limited due to the use of spherical harmonics for specularity. Additionally, we observe that view-dependent spherical harmonics are more prone to overfitting, resulting in flickering artifacts in animation. Please refer to our supplemental video for more details.
In contrast, our approach based on spherical Gaussians is not band-limited and thus achieves high-frequency reflections.

\noindent \textbf{Impact of Data Quality.}
We discover that our approach works even for more relaxed setups such as using a generic template mesh as the base geometry regardless of expressions, and ablating up to 90\% of cameras or 95\% of light patterns from the training data. While we recommend to use our setup to achieve the best quality, this indicates that the proposed method can be applied to much more modest setups. Please refer to Appendix~\ref{sec:ablation} for the experiments.

\section{Conclusion}
\label{sec:conclusion}

We presented Relightable Gaussian Codec Avatars, a novel appearance and geometric representation for relightable 3D head avatars that supports real-time rendering. Our experiments show that high-fidelity relighting of hair, skin, and eyes in all-frequency illuminations is now possible in real-time with the proposed radiance transfer basis composed of spherical harmonics and spherical Gaussians. We have also shown that our choice of the geometric representation based on 3D Gaussian Splatting is critical for strand-accurate hair reconstruction and relighting. Our approach achieves a significant quality improvement in comparison to existing real-time-renderable geometry and appearance models, both qualitatively and quantitatively. 

\noindent \textbf{Limitations and Future Work.}
The current approach requires a coarse mesh and gaze tracking as a preprocessing step, which may be sensitive to tracking failures. Similar to~\cite{yang2023towards}, end-to-end learning together with topology consistent tracking~\cite{li2021topologically,liu2022rapid,bolkart2023instant} is an interesting future work direction to enable scalable training. Extending our approach to in-the-wild inputs also remains a challenge due to the lack of precisely known illumination information. Lastly, rendering a large number of our Gaussian avatars would be difficult, as the relighting operation is performed per individual 3D Gaussian, and scales linearly with the number of avatars. Offloading computation in a per-pixel fragment shader, similar to~\cite{ma2021pixel}, is also an exciting future research. 

{\small
\bibliographystyle{ieee_fullname}
\bibliography{11_references}
}

\clearpage
\appendix
\label{sec:appendix}

\appendix
\section{Network Architecture}
\label{sec:net_arch}

Our head decoder consists of a view-independent decoder and a view-dependent decoder. An expression latent code $\mathbf{z} \in \mathbb{R}^{256}$ is first fed into a single linear layer with a leaky-ReLU, and then reshaped into $256 \times 8 \times 8$. Similarly, the gaze direction of each eye is fed into a linear layer with a leaky-ReLU, and then reshaped into $16 \times 2 \times 2$ for each. The gaze features are then only concatenated where the eye balls are located in the UV space, with the rest zero-padded. 
For view-dependent decoding, we take the unit vector direction from the rendering camera to the head center, and feed it into a linear layer with a leaky-ReLU to obtain a 8-dim latent feature, which is repeated across spatial dimensions for view-conditioning. The input features are concatenated and then fed into both decoders. 
Both the view-independent and view-dependent decoders consist of multiple up-sampling layers based on a transpose convolutional layer ($4 \times 4$ kernel, stride 2) followed by a leaky-ReLU with channel sizes of $(272, 256, 128, 128, 64, 32, 16, 125)$ and $(280, 256, 128, 128, 64, 32, 16, 4)$ respectively.
The eye decoder also uses a similar design while an input spatial resolution to the up-sampling layers of $4 \times 4$. The relative head rotation and position are simply repeated across the spatial dimensions. We also concatenate a visibility mask of eyeballs in UV space by jointly rasterizing the coarse head mesh and the eyeballs to account for the shadows cast by the eyelids. The channel sizes of both view-independent and view-independent layers are $(23, 256, 128, 128, 64, 64, 122)$, $(31, 256, 128, 128, 64, 64, 7)$ respectively.
Note that we use weight normalization~\cite{salimans2016weight} for all linear layers and up-sampling layers, and untied bias~\cite{lombardi2018deep,lombardi2021mixture} for all up-sampling layers.

\section{Discussion: Appearance Representation}
\label{sec:discussion}
In this section, we describe how we derive our specular term from the following rendering equation~\cite{kajiya1986rendering}:
\begin{equation}
\label{eq:render}
\mathbf{c}(\boldsymbol{\omega}_o)=\int_{\mathbb{S}^2} \mathbf{L}(\boldsymbol{\omega}_i) V(\boldsymbol{\omega}_i) \rho(\boldsymbol{\omega}_o, \boldsymbol{\omega}_i) \max (0, \boldsymbol{\omega}_i \cdot \mathbf{n}) \mathrm{d} \boldsymbol{\omega}_i,
\end{equation}
where $\boldsymbol{\omega}_i$ and $\boldsymbol{\omega}_o$ are incoming and outgoing light directions, $\mathbf{L}$ is the incoming light intensity, $V$ is the visibility term, $\rho$ is the BRDF, and $\mathbf{n}$ is the surface normal. Assuming the specular BRDF is represented with the general microfacet model, the specular component of BRDF is defined as follows:
\begin{align}
\rho_S(\boldsymbol{\omega}_o, \boldsymbol{\omega}_i) &=\frac{F(\boldsymbol{\omega}_o, \boldsymbol{\omega}_i) S(\boldsymbol{\omega}_o) S(\boldsymbol{\omega}_i)}{\pi(\boldsymbol{\omega}_i \cdot \mathbf{n})(\boldsymbol{\omega}_o \cdot \mathbf{n})} D(\mathbf{h}) \\
\label{eq:brdf}
&= M(\boldsymbol{\omega}_o, \boldsymbol{\omega}_i) D(\mathbf{h}),
\end{align}
where $F$ is the Fresnel term, $S$ is the geometric attenuation term, and $\mathbf{h}$ is the halfway vector. Following Wang \etal~\cite{wang2009all}, we parameterize the normal distribution function (NDF) $D(\mathbf{h})$ as spherical Gaussian $G_{s}(\mathbf{p}; \mathbf{q}, \sigma)$ (Eq.~6 in the main paper). According to Wang~\etal~\cite{wang2009all}, the remaining term $M$ is smooth and can be approximated as a constant across each Gaussian. After a spherical warping (Eq.~17-22 in~\cite{wang2009all}), we approximate Eq.~\ref{eq:brdf} as:
\begin{equation}
\label{eq:approx}
\rho_S(\boldsymbol{\omega}_o, \boldsymbol{\omega}_i) \approx M(\boldsymbol{\omega}_o, \boldsymbol{\omega}_i)G_{s}(\boldsymbol{\omega}_i; \mathbf{q}, \sigma),
\end{equation}
where $\mathbf{q}$ is the reflection vector. By substituting Eq.~\ref{eq:approx} into Eq.~\ref{eq:render}, our specular term becomes:
\begin{equation}
\label{eq:render_spec}
\int_{\mathbb{S}^2} (V(\boldsymbol{\omega}_i) M(\boldsymbol{\omega}_o, \boldsymbol{\omega}_i)\max (0, \boldsymbol{\omega}_i \cdot \mathbf{n}))\mathbf{L}(\boldsymbol{\omega}_i)G_{s}(\boldsymbol{\omega}_i; \mathbf{q}, \sigma)  \mathrm{d} \boldsymbol{\omega}_i.
\end{equation}
When $\sigma \ll 1$, the value inside the integral is $0$ unless $\boldsymbol{\omega}_i$ is close to $\mathbf{q}$, which is determined by the input view $\boldsymbol{\omega}_o$. Therefore, we further approximate Eq.~\ref{eq:render_spec} by moving and combing all view-dependent terms together (denoted as $v_k$) except the incoming radiance $\mathbf{L}$ and NDF $G_s$ as follows:
\begin{equation}
\label{eq:final_spec}
\mathbf{c}^{\textrm{specular}}_{k} = v_k(\boldsymbol{\omega}_o) \int_{\mathbb{S}^2} \mathbf{L}(\boldsymbol{\omega}_i)G_s(\boldsymbol{\omega}_i; \mathbf{q}, \sigma) \mathrm{d} \boldsymbol{\omega}_i.
\end{equation}
Importantly, we parameterize $v_k(\boldsymbol{\omega}_o)$ using a neural network, enabling end-to-end optimization with the remaining components to faithfully reproduce image observations. Thus, our model is flexible enough to represent specular reflection beyond the general microfacet model~\cite{wang2009all} or single-bounce reflection.
We empirically find that this simple formulation is fast to compute, and stable to optimize. It also supports modeling both diffuse and highly reflective areas in a unified manner. 
In our paper, we constrain the specular BRDF to monochrome to prevent the specular term from overfitting diffuse components. 
Supporting color changes in specular highlights caused by dielectric materials or multi-bounce specular reflection can be addressed in future work.

\section{Ablation Study}
\label{sec:ablation}

In this section, we provide ablation studies to validate our key design choices.

\begin{figure}[!ht]
\begin{center}
\includegraphics[width=\linewidth]{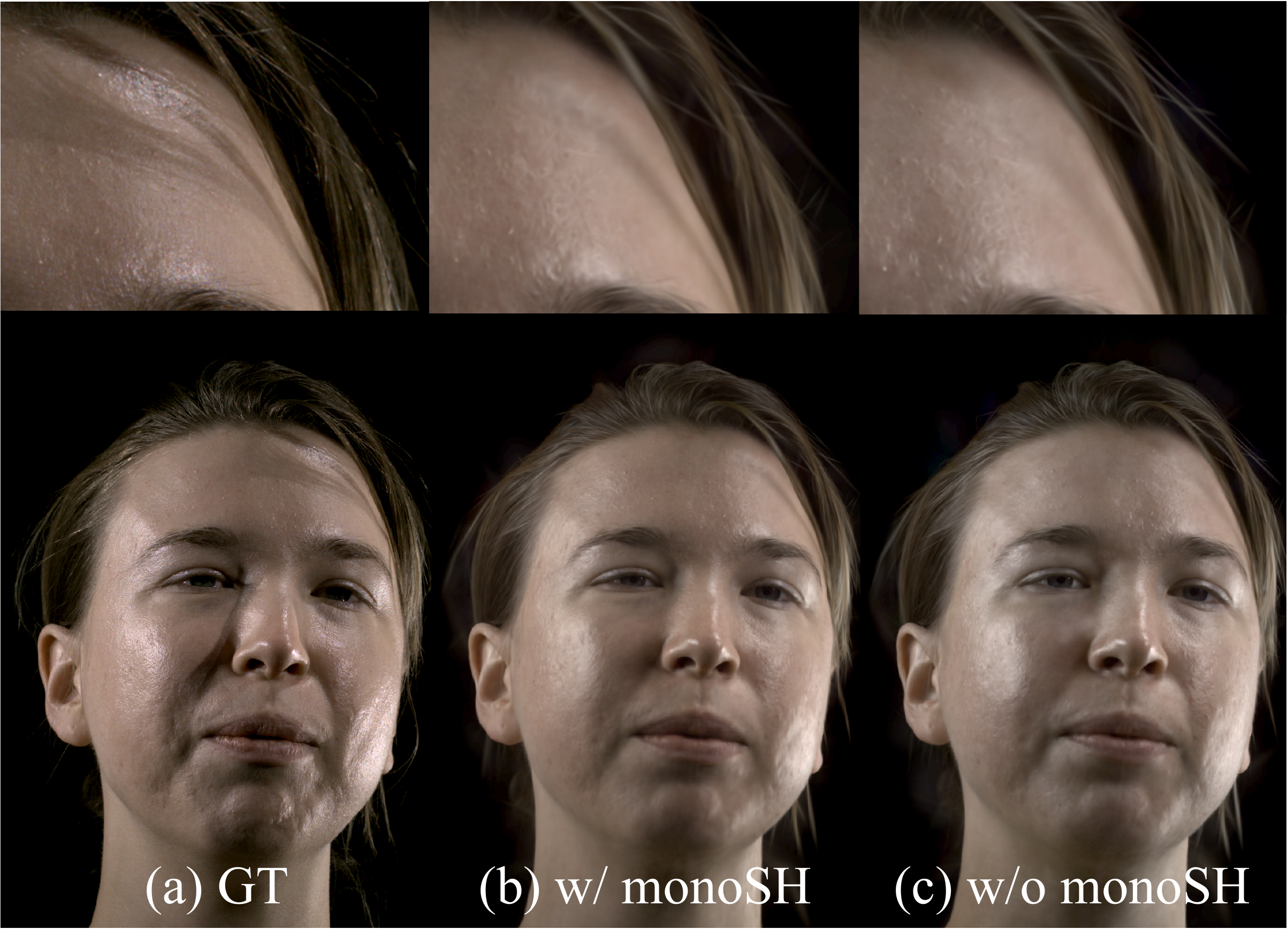} 
\caption{\textbf{Ablation Study: Monochrome SH}. Compared to a held out frame (a), using higher-order monochrome SH coefficients (b) improves the sharpness of shadows compared to a model without them (c).}
\label{fig:monosh}
\vspace{-15pt}
\end{center}
\end{figure}

\noindent \textbf{Higher-order Monochrome SH.}
Our diffuse color is based on spherical harmonics. To support high-frequency shadows, our model decodes additional monochrome SH coefficients up to 8-th order. We compare our approach with one where we remove 4-th to 8-th order monochrome SH coefficients with the remaining components being identical. 
Fig.~\ref{fig:monosh} shows that our approach captures more precise shadows. The quantitative evaluation in Tab.~\ref{tab:ablation} also shows that adding the monochrome SH coefficients improves overall reconstruction accuracy. Note that while some recent works utilize explicitly computed shadow maps~\cite{bi2021deep,iwase2023relightablehands,sarkar2023litnerf}, this is intractable for real-time relighting with high-frequency environments. Improving the sharpness of shadows in real-time relighting even further is an interesting direction for future work.

\begin{table}[t]
    \centering
    \begin{footnotesize}
    \caption{\textbf{Ablation Study}. The top three techniques are highlighted in \textcolor{rred}{red}, \textcolor{oorange}{orange}, and \textcolor{yyellow}{yellow}, respectively. We use 3D Gaussians with the explicit eye models for the geometric representations.}
    \vspace{-5pt}
    \label{tab:ablation}
    \setlength{\tabcolsep}{2mm}{
    \renewcommand\arraystretch{0.975}
    \resizebox{1.0\columnwidth}{!}{
    \begin{tabular}{c|ccc}
    \toprule
     \multirow{2}{*}{Method}& \multicolumn{3}{c}
    {Metrics}\\
    & \footnotesize{PSNR\;$\uparrow$} & \footnotesize{SSIM\;$\uparrow$} & 
    \footnotesize{LPIPS\;$\downarrow$} \\
    \midrule
    Ours &\cellcolor{rred}34.042&\cellcolor{oorange}0.858&\cellcolor{oorange}0.148\\
    Ours w/o monoSH&33.762&0.853&0.152\\
    Ours w/o view-dep nml.&\cellcolor{oorange}33.927&\cellcolor{rred}0.864&\cellcolor{yyellow}0.148\\
    SG~\cite{wang2009all,zhang2021physg}&\cellcolor{yyellow}33.778&\cellcolor{yyellow}0.855&\cellcolor{rred}0.147\\
    \bottomrule
    \end{tabular}}}
    \vspace{-5pt}
    \end{footnotesize}
\end{table}

\noindent \textbf{View-dependent Normal.}
Another component in our appearance model is the view-conditioned surface normal. We compare our approach with one where we remove view-conditioning when decoding the surface normal. Interestingly, the improvement does not clearly appear in both qualitative and quantitative comparisons (see Tab.~\ref{tab:ablation}). We hypothesize that our view-conditioned visibility term can compensate for some of the errors caused by view-independent surface normals in cylindrical regions. While this allows the baseline using view-independent normals to achieve comparable performance under discrete point lights, this would likely cause inaccurate reflection on continuous environments. 
We keep our view-conditioned normals as this offers a more geometrically correct interpretation for the cylinder-like 3D Gaussians.

\begin{figure}[!ht]
\begin{center}
\includegraphics[width=\linewidth]{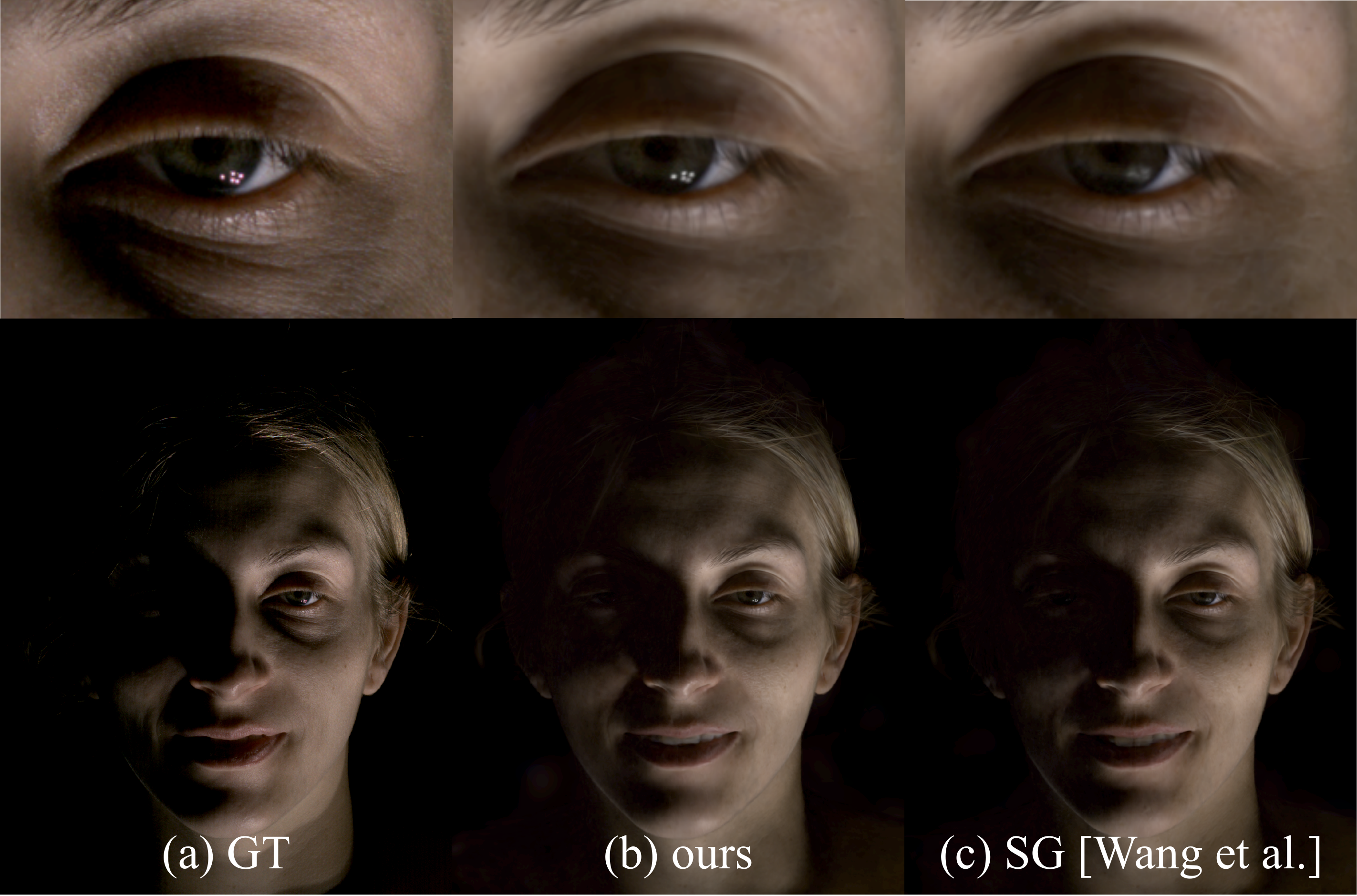} 
\caption{\textbf{Ablation Study: Spherical Gaussian Representation}. Compared to a held out frame (a), our angle-based SG formulation (b) leads to more accurate recovery of eye glints than the conventional cosine-based SG formulation~\cite{wang2009all} (c).}
\label{fig:sg}
\vspace{-15pt}
\end{center}
\end{figure}

\noindent \textbf{Spherical Gaussian Formulation.}
Prior works using spherical Gaussians~\cite{wang2009all,zhang2021physg} typically use a different parametrization $G(\mathbf{p} ; \mathbf{q}, \lambda, \mu)=\mu e^{\lambda(\mathbf{p} \cdot \mathbf{q}-1)}$.
We compare our method with this formulation of spherical Gaussians with the remaining parts being identical. 
While the overall results are comparable quantitatively, Fig.~\ref{fig:sg} shows that our parameterization better captures sharp eye glints, which is critical for accurate all-frequency reflections.

\begin{figure}[!ht]
\begin{center}
    \includegraphics[width=1.0\linewidth]{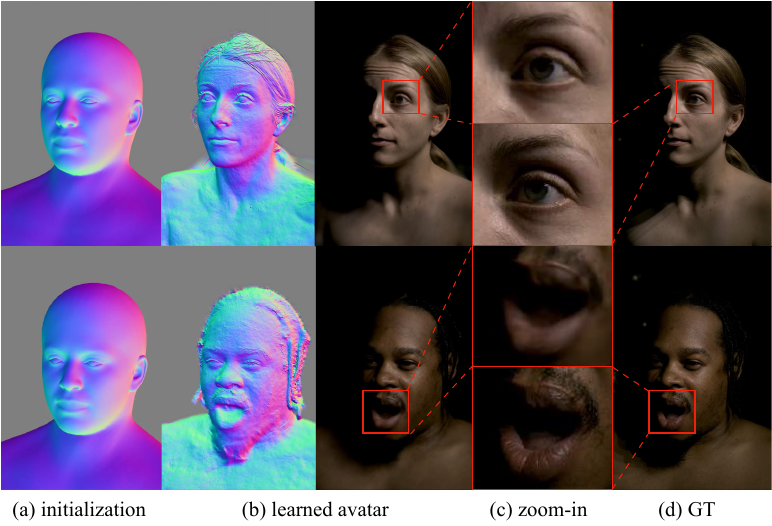}
    \vspace{-20pt}
    \caption{\textbf{Ablation Study: Only Rigid Tracking}. We use a generic head template as the base mesh regardless of facial expressions (a). Compared to GT (d), our model with only rigid head pose tracking and a generic template achieves surprisingly good reconstruction (b, c).}
    \label{fig:temp_track}
\vspace{-15pt}
\end{center}
\end{figure}

\noindent \textbf{Person-specific mesh and non-rigid tracking required?}
We train our model with a generic head template as initialization regardless of facial expressions (Fig.~\ref{fig:temp_track} (a)). We also disable the geometry loss $\mathcal{L}_{\mathrm{geo}}$ such that the positions of Gaussians are only updated through differentiable rendering. In other words, we use only the estimated rigid headpose and gaze directions as input. Although slightly worse registration sometimes leads to lack of eye glints and blurrier extreme facial expressions, the model achieves surprisingly good reconstruction as shown in Fig.~\ref{fig:temp_track} (b). This indicates that our Gaussian-based representation is flexible enough to register even if the initialization is poor. The dependency on accurate non-rigid surface tracking can be optionally removed at the risk of slight quality degradation (e.g., lack of eye reflections).

\begin{figure}[!ht]
\begin{center}
    \includegraphics[width=1.0\linewidth]{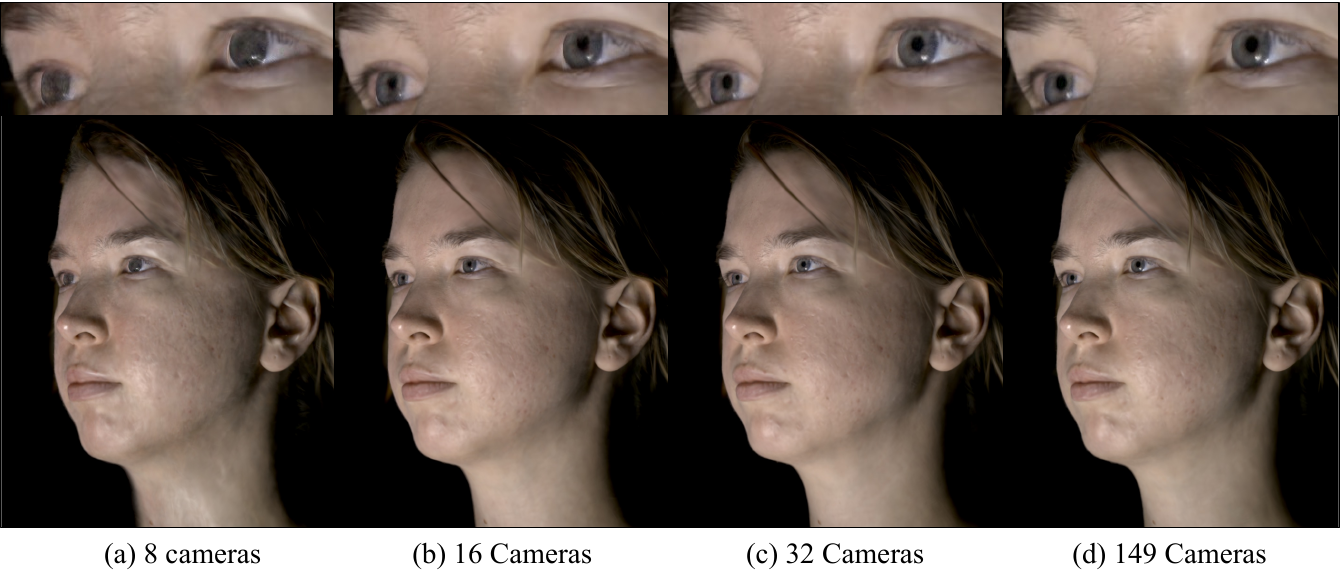}
    \vspace{-20pt}
    \caption{\textbf{Ablation Study: Number of cameras for decoder training}. We vary the number of cameras used for rendering supervision (a) 8 cameras, (b) 16 cameras, (c) 32 cameras, (d) the full 149 cameras. We show results of novel view generation on a training frame.}
    \label{fig:cam_ablation}
\vspace{-15pt}
\end{center}
\end{figure}

\noindent \textbf{Effect of the number of cameras.}
We train our decoder model with varying numbers of cameras to analyze the sensitivity of the method to capture setup specifics, and show results of novel view synthesis on a training frame (Fig.~\ref{fig:cam_ablation}). Using as few as 32 cameras seems to yield good results, with 8 cameras showing noticeably degraded quality, and 16 cameras showing some artifacts, especially in the eyes. Conversely, using more than 32 cameras yields diminishing returns. We hypothesize that higher capacity modeling would be required to fully utilize the available data. (Note also that any rigid head motion present across the training frames creates additional virtual viewpoints---training on a single frame would yield much worse results). 

\begin{figure}[!ht]
\begin{center}
    \includegraphics[width=1.0\linewidth]{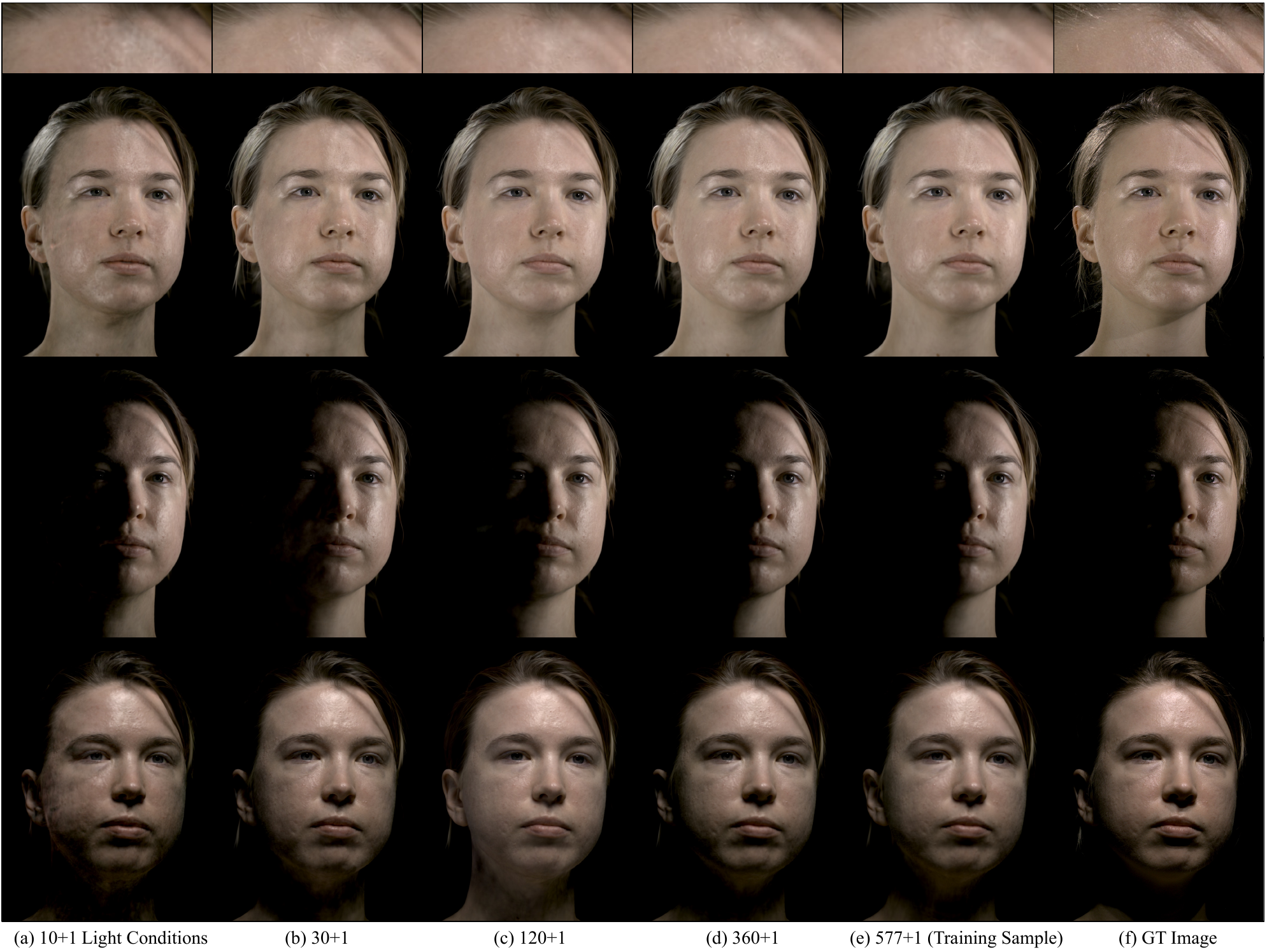}
    \vspace{-20pt}
    \caption{\textbf{Ablation Study: Number of light conditions used in training.}. We vary the number of light conditions used for rendering supervision (a) 10+1 (10 partial illuminations and 1 uniform illumination), (b) 30+1, (c) 120+1, (d) 360+1, (e) the full set of illuminations (including the test sample), and (f) the ground truth image. We show results on held out illuminations for a training frame and camera.}
    \label{fig:light_ablation}
\vspace{-15pt}
\end{center}
\end{figure}

\noindent \textbf{Effect of the number of lighting conditions.}
We train our decoder model with varying numbers of light conditions and show an unseen light condition on a training frame (Fig.~\ref{fig:light_ablation}). We note two limitations of this study: (1) because we use temporal multiplexing, the comparisons use different numbers of training frames (as all frames from other light conditions need to be discarded), and (2) we cannot hold out physical lights as our light conditions trigger multiple lights simultaneously. However, the results show that using even 10\% to 20\% percent light conditions can yield acceptable results, potentially again limited by capacity and learning variance.

\section{Performance}

For all identities, we use $1024 {\times} 1024 = 1$~Mi Gaussians for the evaluation and results on the paper, and $512 {\times} 512 = 256$~Ki Gaussians for the VR demo shown in the video. We observe that increasing the number of Gaussians leads to quality improvement at the cost of slower decoding and rendering. The $1024^2$ model takes 12.84 ms for splatting, and the $512^2$ model takes 6.40 ms for splatting on NVIDIA A100. We use $512^2$ for the VR demo to improve the framerate. We do not apply any pruning of Gaussians.
Tab.~\ref{tab:perf} shows the inference time of each method. All Gaussian-based models including ours converge within 3 days and MVP-based models require twice as many
032 iterations (400 K) for convergence.

\begin{table}[t!]
    \centering
    \begin{footnotesize}
    \caption{\textbf{Performance of each method}.}
    \vspace{-5pt}
    \label{tab:perf}
    \setlength{\tabcolsep}{2mm}{
    \renewcommand\arraystretch{0.975}
    \resizebox{1.0\columnwidth}{!}{
    \begin{tabular}{c|c|c|c}
    \toprule
     &Geometry & Appearance& Inference (ms) \\
    \midrule
    A & \multirow{2}{*}{Ours w/ EEM} &EyeNeRF~\cite{li2022eyenerf}&35\\
    B & &Ours&31\\
    \midrule
    C & \multirow{3}{*}{Ours} &EyeNeRF~\cite{li2022eyenerf}&20\\
    D & &Linear~\cite{yang2023towards}&6\\
    E& &Ours&18\\
     \midrule
    F & \multirow{3}{*}{MVP~\cite{lombardi2021mixture}} &EyeNeRF~\cite{li2022eyenerf}&43\\
    G& &Linear~\cite{yang2023towards}&6\\
    H & &Ours &34\\
    \bottomrule
    \end{tabular}}}
    \vspace{-10pt}
    \end{footnotesize}
\end{table}

\section{Ethical Concerns}
\label{sec:ethics}
Our model is only applied to a few consenting subjects captured in a dense multiview capture system. In addition, the expression latent space is personalized for each individual to capture subtle expressions.
These effectively limit the use case to driving ones' own avatars only with their consent.

\end{document}